\documentclass[10pt,twocolumn,letterpaper]{article}

\usepackage{ijcb2023/cvpr}
\usepackage{times}
\usepackage{epsfig}
\usepackage{graphicx}
\usepackage{amsmath}
\usepackage{amssymb}
\usepackage{enumerate}
\usepackage[boxed,linesnumbered]{algorithm2e}
\SetAlFnt{\footnotesize}
\SetAlCapNameFnt{\small}
\SetAlCapFnt{\small}
\usepackage{xcolor}
\usepackage[hyphens]{url}
\usepackage{multirow}
\usepackage{subcaption}
\usepackage{caption}
\usepackage{svg}
\usepackage[accsupp]{axessibility}



\makeatletter
\def\thanks#1{\protected@xdef\@thanks{\@thanks
        \protect\footnotetext{#1}}}
\makeatother

\begin{document}

\title{Two-Dimensional Dynamic Fusion for Continuous Authentication\thanks{\scriptsize
  Copyright~\copyright~2023 IEEE. Personal use of this material is permitted. Permission from IEEE must be obtained for all other uses, in any current or future media, including reprinting/republishing this material for advertising or promotional purposes, creating new collective works, for resale or redistribution to servers or lists, or reuse of any copyrighted component of this work in other works.}}

\author{Nuttapong Attrapadung\footnotemark[1]{},\, Goichiro Hanaoka\footnote[1]{},\, Haochen M.~Kotoi-Xie\footnote[2]{},\, Takahiro Matsuda\footnote[1]{},\\ Takumi Moriyama\footnote[2]{},\, Takao Murakami\footnote[3]{},\, Hidenori Nakamura\footnote[2]{},\, Jacob C.~N.~Schuldt\footnote[1]{},\\ Masaaki Tokuyama\footnote[2]{},\, Jing Zhang\footnote[2]{}\\
\vspace{-0.2cm}\\
\footnote[1]{}\, National Institute of Advanced Industrial Science and Technology (AIST), Japan\vspace{-0.1cm}\\
{\tt\small \{\,n.attrapadung, hanaoka-goichiro, t-matsuda, jacob.schuldt\,\}@aist.go.jp}\vspace{0.2cm}\\
\footnote[2]{}\, AnchorZ Inc., Japan\vspace{-0.1cm}\\
{\tt\small \{\,kotoi, moriyama, nakamura, tokuyama, j-zhang\,\}@anchorz.co.jp}\vspace{0.2cm}\\
\footnote[3]{}\, Institute of Statistical Mathematics (ISM), Japan\vspace{-0.1cm}\\
{\tt\small 	tmura@ism.ac.jp}}

\maketitle

\thispagestyle{empty}

\begin{abstract}
Continuous authentication has been widely studied to provide high security and usability for mobile devices by continuously monitoring and authenticating users. 
Recent studies adopt multibiometric fusion for continuous authentication to provide high accuracy even when some of captured biometric data are of a low quality. 
However, existing continuous fusion approaches are resource-heavy as they rely on all classifiers being activated all the time and may not be suitable for mobile devices.

In this paper, we propose a new approach to multibiometric continuous authentication: 
two-dimensional dynamic fusion. 
Our key insight is that multibiometric continuous authentication calculates two-dimensional matching scores over classifiers and over time. 
Based on this, we dynamically select a set of classifiers based on the context in which authentication is taking place, and fuse matching scores by multi-classifier fusion and multi-sample fusion. 
Through experimental evaluation, we show that our approach provides a better balance between resource usage and accuracy than the existing fusion methods.
In particular, we show that
our approach provides higher accuracy than the existing methods with the same number of score calculations by adopting multi-sample fusion.
\end{abstract}


\section{Introduction}

Mobile devices such as smartphones are ubiquitous in today's society, and with their increased use, these devices store a large amount of sensitive user data, including photos, passwords, 
purchase history, banking, and even payment information~\cite{enterpriseappstoday}.
While banking and payment information must be protected for obvious reasons, the exposure of other types of personal information can likewise have serious consequences; an IDG Research survey~\cite{lookout} 
estimated that in the past, a significant fraction of smartphone thefts led to identity theft.  
This has made malicious smartphone access a significant security risk.
However, at the same time, secure user authentication remains a challenging task; the average smartphone user checks his smartphone $96$ times a day~\cite{foxnews}, which makes it impractical to enter high-entropy passwords or 
use a different device for two-factor authentication.
Thus, any smartphone authentication mechanism must be efficient and unobtrusive.
Moreover, smartphone 
users frequently unlock their phones in potentially malicious environments, and might leave the smartphone unguarded, e.g.,\ on a table or in a bag 
from which the phone can easily be stolen.
Thus, an authentication model developed for desktop users (i.e., unlock once at the beginning of a session, and lock once the session is done) leaves smartphones vulnerable to attacks while 
in an unlocked state. 

Continuous authentication~\cite{Reiner11} is based on a fundamentally different authentication model in which the user is continuously authenticated via physical and behavioral traits, 
including direct biometric authentication mechanisms such as facial and voice recognition, but also soft biometrics such as touch patterns, gait, motion, and location information.
Modern smartphones feature a wide range of sensors that allow the capture of such input and have furthermore become sufficiently powerful to process this input.
This raises the prospect of restructuring user authentication on smartphones based on continuous authentication, potentially bringing both usability and security advantages.

The ideal continuous authentication system seamlessly and automatically authenticates the user once he starts using the device, 
but immediately locks the device once a different user attempts to interact with this. 
However, realizing a system coming close to this is a challenging task.
When the user is not explicitly engaging with an authentication mechanism, the captured input could be of lower quality. 
For example, 
the user might position his head such that the view of his face from the camera is obstructed. 
For another example, 
the user might move into a poorly lit area, which will make face recognition challenging.
Immediately locking the device when one authentication mechanism fails would unnecessarily obstruct the user workflow, whereas the aim is to maintain a seamless usability experience.

To address this, continuous authentication is often proposed as a 
multibiometric system \cite{CHCBJ15,KPS16,SivRagSimZic18,SR19} that fuses multiple sources of biometric information
to identify a user.
For example, Sivasankaran \emph{et al.}~\cite{SivRagSimZic18} use the context in which authentication is taking place (i.e., the environment, such as light intensity and noise level) for fusion in continuous authentication. 
Specifically, they propose an approach denoted \emph{context-weighted majority algorithm} (CWMA), which optimizes parallel fusion of multiple classifiers (e.g., two face matchers and one voice matcher) by applying context-dependent weights when fusing authentication results. 
Relying on the fused output of several classifiers is often more secure and reliable than any of the individual mechanisms on their own~\cite{RutGab00}.
However, this comes with the cost of having to activate multiple classifiers and process the output of these, which will occupy resources on the device and increase power consumption.
In particular, the latter can be critical on smartphones, which are often optimized to reduce power consumption in order to increase the time the device can be used without having to recharge. 
This issue is exacerbated by the fact that the continuous authentication system should ideally be authenticating the user constantly to ensure that the device is locked as soon as a different user tries to operate this.
Finding the optimal approach to implementing continuous authentication that balances resource usage, usability, and security remains an open question. 

\smallskip
\noindent{\textbf{Our Contributions.}}~~We propose a new approach to 
multibiometric fusion in 
continuous authentication, which we call \textit{two-dimensional dynamic fusion}. 
The key insight underlying our approach is that 
multibiometric continuous authentication can be viewed as a fusion of \textit{two-dimensional} matching scores (similarities or distances) over classifiers and over time. 
We dynamically determine which scores to calculate in the two-dimensional space 
based on the context (i.e.,\ the environment). 
Here, we use the context because it is a good predictor of the classifiers most likely to successfully authenticate the user; e.g., we should use facial (resp.~voice) recognition in a noisy (resp.~low light) environment.
In addition, the context can often be determined via sensors that require very little computational resources (such as light sensors or microphones) and which can be running in the background continuously polling the environment at regular intervals.  

The advantage of our approach over state-of-the-art context-based fusion such as CWMA \cite{SivRagSimZic18} lies in the balance of resource usage and security which ultimately leads to improved usability. 
As explained above, CWMA calculates all the scores, which 
is resource-heavy and not ideal for continuous authentication on mobile devices where frequent authentication is a desirable feature.
In contrast, our approach uses the context to determine a set of classifiers that has a sufficiently high combined expected success probability of authenticating a genuine user.
By only activating this set of classifiers, resource usage can be minimized.

In other words, the fact that we fuse two-dimensional scores implies that two types of multibiometric fusion are possible: multi-classifier fusion\footnote{Following \cite{SivRagSimZic18}, we use the term ``multi-classifier.'' 
Since a classifier consists of a sensor and matching algorithm for a specific modality, multi-classifier includes multi-modal, multi-sensor, and multi-algorithm \cite{Ross06}.} and multi-sample fusion (i.e., a fusion of multiple scores from the same classifier) \cite{Ross06}. 
Our approach (implicitly) selects \textit{the type of multibiometric fusion} (i.e., multi-classifier, multi-sample, or the combination of the two) to maximize the accuracy under the constraint of limited computational resources.
For example, in a noisy environment, the fusion of two face samples (e.g., multi-sample) could provide higher accuracy than the fusion of face and voice samples obtained at the same time (e.g., multi-classifier). 
In fact, we show that our approach provides high accuracy by adopting multi-sample fusion in a context where 
some classifiers provide 
poor performance. 

In summary, our main contributions are as follows: 
\begin{itemize}
    \item We propose two-dimensional dynamic fusion for continuous authentication, which continuously selects a set of classifiers based on the context information to maximize the accuracy with limited resource usage. 
    \item Through experimental evaluation, we show that our approach provides a good balance between resource usage and accuracy. 
    In particular, we show that our multi-sample strategy is effective. 
    Specifically, we compare our approach with the existing fusion methods (e.g., CWMA \cite{SivRagSimZic18}, max/sum rule after z-score normalization \cite{zscore}) applied to a certain time instant. 
    Our experimental results show that our approach can achieve almost the same accuracy as the existing methods with fewer score calculations and higher accuracy with the same number of score calculations.
\end{itemize}

\smallskip
\noindent{\textbf{Related Works.}}~~Continuous authentication \cite{CHCBJ15,KPKPS22,KPS16,RS22,SivRagSimZic18,SR19} has been recently studied to continuously authenticate users based on biometric data. 
Some of the previous works adopt multibiometric fusion \cite{Ross06} for continuous authentication. 
For example, Crouse \etal \cite{CHCBJ15} correct the uprightness of face images using a gyroscope, accelerometer, and magnetometer data. 
Then they fuse the uprightness-corrected face images in a session. 
Kumar \etal \cite{KPS16} combine typing patterns, swiping gestures, and phone movement patterns for smartphones. 
Smith-Creasey and Rajarajan \cite{SR19} propose a fusion algorithm based on the Dempster-Shafer theory to incorporate uncertainty into matching scores. 
None of them consider the context (e.g., sitting in a dark room, talking in a noisy environment) in their fusion algorithms. 

Sivasankaran \etal \cite{SivRagSimZic18} propose a context-aware fusion algorithm called CWMA, which trains a weight for each classifier and each context and calculates a final score as a weighted sum of scores. 
They show that context information is useful for improving the accuracy of continuous authentication. 
However, CWMA always uses all classifiers and is resource-heavy, as explained above. 
To address this issue, we propose two-dimensional dynamic fusion, 
which selects classifiers based on the context. 

Another related approach is quality-based classifier switching \cite{BBSVN15,VSNR10}\footnote{This approach is called quality-based context switching in \cite{BBSVN15,VSNR10}. 
Here, they use ``context'' to mean classifiers (e.g., modalities, matchers). 
We use the term ``classifier'' to avoid confusion.}, which selects some classifiers in biometric fusion based on the quality (e.g., brightness, energy spectrum) of biometric samples. 
Our approach differs from their approach in that ours 
fuse two-dimensional scores over classifiers and over time and 
(implicitly) selects \textit{the type of multibiometric fusion}, such as multi-modal and multi-sample, as explained above. 
In particular, our multi-sample strategy is effective, as shown in our experiments.

\begin{figure}[t]
	\centering
	\includegraphics[scale=0.3]{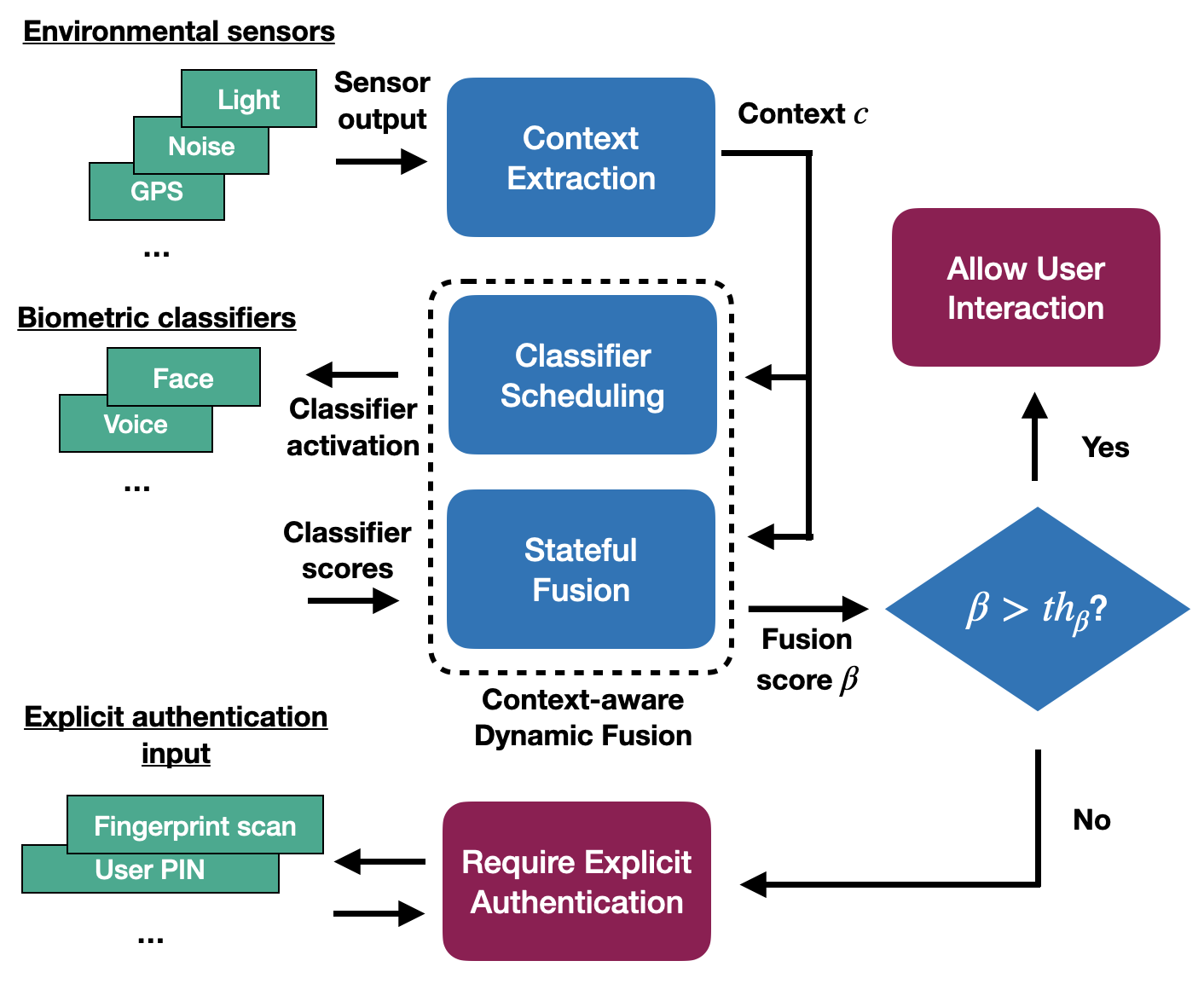}
        \vspace{-2mm}
	\caption{Overview of 
        our continuous authentication framework.
        }\label{fig:framework-overview}
\vspace{2mm}
	\centering
	\includegraphics[scale=0.21]{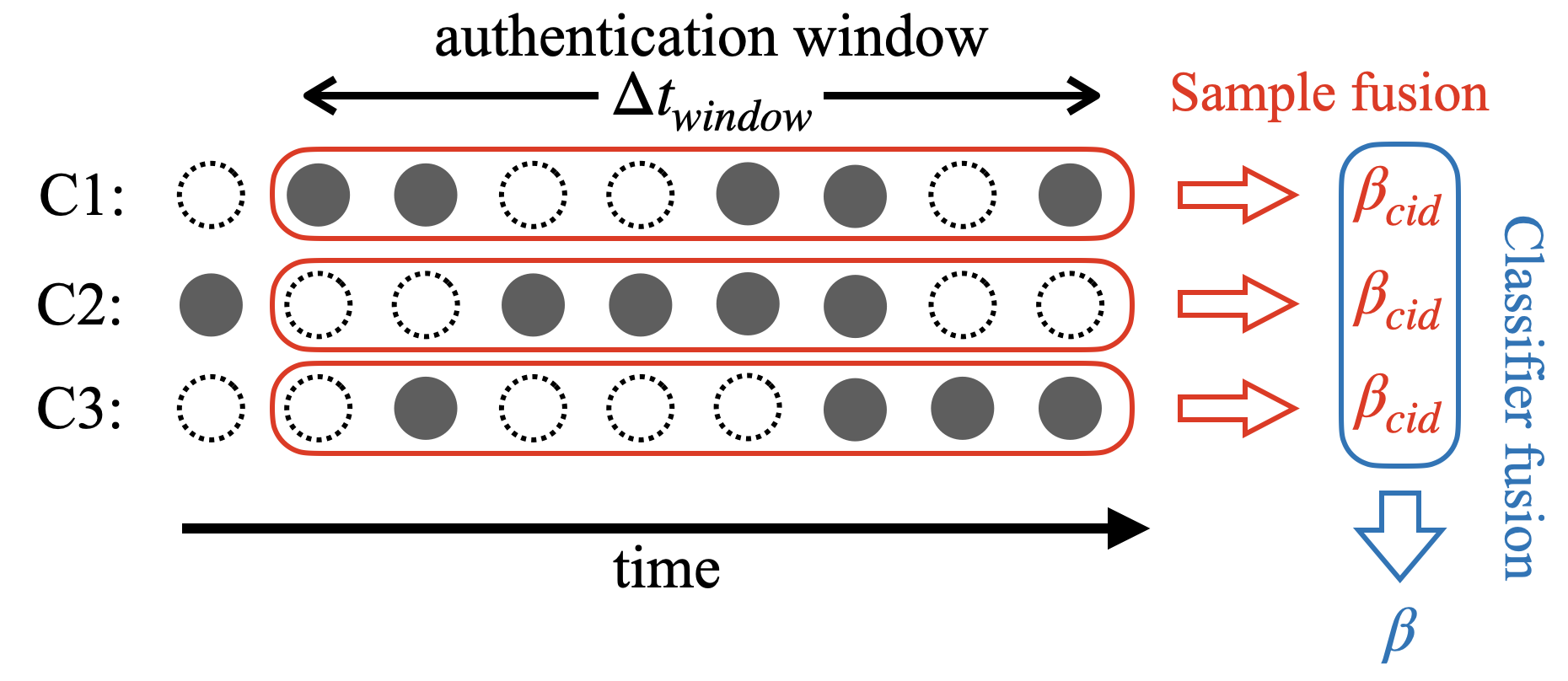}
        \vspace{-2mm}
	\caption{Overview of our two-dimensional dynamic fusion approach for three classifiers (C1, C2, C3). Solid circles indicate available classification scores.}\label{fig:fusion}
\end{figure}

\section{Proposed Continuous Authentication}
In this section, we propose our continuous authentication. 
Section~\ref{sec:overview} provides an overview. 
Sections~\ref{sec:sensor-scheduling} and \ref{sec:fusion} explain 
our main components: classifier activation scheduling and stateful fusion. 
Section~\ref{sec:main_loop} shows the main authentication loop in our continuous authentication. 

\subsection{Overview}
\label{sec:overview}
An overview of the proposed continuous authentication framework is given in Figure~\ref{fig:framework-overview}.
At the core of the framework are a classifier scheduling algorithm and a corresponding fusion algorithm 
that, in combination, implements our context-aware dynamic multibiometric fusion.
The classifier scheduling algorithm will dynamically activate classifiers based on probability estimates of each classifier successfully authenticating a genuine user in the given context determined via environmental sensor outputs\footnote{These environmental sensors should measure environmental variables relevant to estimating the efficacy of the used biometric classifiers and should be sufficiently lightweight to be continuously running in the background. In practice, these will be, e.g.,\ a light sensor detecting the level of light, a microphone detecting the noise level, and similar types of sensors.}. 
The set of classifiers to be activated is chosen as the set which simultaneously has a sufficiently high combined authentication probability and the lowest recourse cost (see Section~\ref{sec:sensor-scheduling} for a detailed description).
The corresponding fusion algorithm is based on the \emph{history} of obtained classifier scores and will compute a fusion score $\beta$ based on the classifier scores obtained within a given (context-dependent) time frame which we denote the \emph{authentication window}. 
In other words, the fusion algorithm implements a two-dimensional fusion approach combining multi-modal and multi-sample fusion as illustrated in Figure~\ref{fig:fusion}. 
Note that the classifier authentication scores available to the fusion algorithm are dynamically decided by the classifier scheduling algorithm, and hence, the classifier scheduling algorithm implicitly decides the type of fusion done in the fusion algorithm.
The fusion algorithm is described in Section~\ref{sec:fusion}.

Finally, based on a comparison between the fusion score $\beta$ and an authentication threshold $th_\beta$,
the device is either kept in an unlocked state or locked; i.e.,\ the user is required to authenticate via an explicit authentication method (e.g., pin code, fingerprint scan, etc.).
In the following subsections, the details of the framework will be described.

\subsection{Context-Aware Classifier Scheduling}
\label{sec:sensor-scheduling}
The classifier scheduling algorithm will decide upon the best possible set of classifiers to activate based on an \emph{a priori} probability estimate for successful authentication (true acceptance rate) for each classifier in a given context (e.g., the performance of facial recognition in different levels of light), and is parameterized by a probability threshold $th_p$.
Furthermore, the algorithm is assumed to have access to the resource cost of activating a set of classifiers and the time required to obtain classifier scores (i.e.,\ input capture and processing).
This is captured via the following algorithms:
\begin{description}
	\item[$\mathtt{AuthProb}(cid,c)$]: On input a classifier id $cid$ and context $c$, this algorithm returns an estimated true acceptance rate. 
    \vspace{-2mm}
	\item[$\mathtt{Time}(cid)$]: On input a 
   classifier id $cid$, 
    this algorithm returns the time required to capture input from 
    the classifier $cid$ 
    and process this.
    \vspace{-2mm}
	\item[$\mathtt{Cost}(\mathcal{S})$]: On input a set of classifier ids $\mathcal{S}$, this algorithm returns a scalar indicating the total resource cost of capturing input from 
    $\mathcal{S}$ and processing this.
\end{description}   

To estimate the probability of successful authentication using a set $\mathcal{S}$ of classifiers, the activation scheduling algorithm heuristically assumes independence of classifier authentication success. 
More precisely, the set $\mathcal{S}$ is considered a valid candidate set if
\[
	th_p \; \leq \; 1 - \prod_{cid \in \mathcal{S}} (1 - \mathtt{AuthProb}(cid,c))  
\]  
i.e., if the combined a priori probability of successful authentication is above the given threshold $th_p$.
The activation scheduling algorithm then simply selects the candidate set with the combined lowest resource cost as the set of classifiers to be activated.

As a fail-safe mechanism, the classifier scheduling algorithm will ensure that all classifiers have been activated at least once in an attempt to authenticate the user before a potential lock of the device.
This is achieved by scheduling any classifier which would otherwise not have time to complete an authentication attempt.
Specifically, let $t_{crit}$ be the \emph{critical} time at which the device is locked if no additional authentication input is obtained, 
$t_{now}$ be the current time, 
$\Delta t_{crit}$ be the time period until $t_{crit}$, i.e.,\ $\Delta t_{crit} = t_{crit} - t_{now}$, and $\Delta t_{delay}$ be the time between each invocation of the classifier scheduling algorithm (see Section~\ref{sec:main_loop} for how $t_{crit}$ and $\Delta t_{crit}$ are derived)\footnote{We use the notation $t$ to denote a time \emph{instant} and $\Delta t$ to denote a time \emph{period/interval}.}.
Then a classifier $cid$ is activated if $\Delta t_{cid} + \Delta t_{delay} \geq \Delta t_{crit}$, where $\Delta t_{cid} \gets \mathtt{Time}(cid)$, regardless of the cost.

The classifier scheduling algorithm is shown as Algorithm~\ref{alg:sensor_activation}.
Note that if no valid candidate set is found, the classifier scheduling algorithm will default to activating all classifiers.

\SetKwInput{KwPar}{Parameters}
\begin{algorithm}
\KwIn{$(\mathcal{S},c,\Delta t_{crit})$ where\\ 
~~~$\mathcal{S}=\{cid\}$: classifier ids;\\ 
~~~$c$: context;\\ 
~~~$\Delta t_{crit}$: time period until the critical time;}
\KwPar{$(th_p,\Delta t_{delay})$ where\\
~~~$th_p$: classifier probability threshold;\\ 
~~~$\Delta t_{delay}$: delay between invocations.}
\KwOut{Set $\mathcal{S}_{act}$ of classifier ids to be activated.}
\BlankLine

\tcc{Time-critical classifiers}
$\mathcal{S}_{crit} \gets \emptyset$\;
\ForEach{$cid \in \mathcal{S}$}{
	$\Delta t_{cid} \gets \mathtt{Time}(cid)$\;
	\If{$\Delta t_{cid} + \Delta t_{delay} \geq \Delta t_{crit}$}{$\mathcal{S}_{crit} \gets \mathcal{S}_{crit} \cup \{cid\}$\;}
}

\BlankLine
\tcc{Compute prob.\ and cost for candidate sets}
$\mathcal{S}_{cand} \gets \emptyset$\; 
\ForEach{subset $\mathcal{S}_{sub} \subseteq \mathcal{S}$}{
        $p_{cand} \gets 1 - \prod_{cid \in \mathcal{S}_{sub}} (1 - \mathtt{AuthProb}(cid,c))$\;
	$c_{cand} \gets \mathtt{Cost}(\mathcal{S}_{sub} \cup \mathcal{S}_{crit})$\;
	\If{$p_{cand} > th_p$}{$\mathcal{S}_{cand} \gets \mathcal{S}_{cand} \cup \{ (c_{cand}, \mathcal{S}_{sub} \cup \mathcal{S}_{crit}) \}$}
}
\BlankLine

\If{$\mathcal{S}_{cand} = \emptyset$\tcc*{No candidate set found}}{\Return $\mathcal{S}$}

\BlankLine
\tcc{Find $(c_{cand},\mathcal{S}_{act}) \in \mathcal{S}_{cand}$ with minimum resource cost $c_{cand}$}
$(c_{cand},\mathcal{S}_{act}) \gets \mathtt{Min}(\mathcal{S}_{cand})$\;
\Return $\mathcal{S}_{act}$\;

\caption{Classifier scheduling algorithm $\mathtt{ClassifierSchedule}$.}\label{alg:sensor_activation}
\end{algorithm}

\subsection{Stateful Fusion}\label{sec:fusion}

Our fusion approach is based on the notion of an \emph{authentication window} $\Delta t_{window}$ within which a user must be successfully authenticated to be allowed continued use of the mobile device.
The authentication window itself is derived from the current context\footnote{Letting the authentication window be context-dependent allows the continuous authentication system to implement a more flexible security policy, e.g., by letting the window length depend on whether the device is in a low-risk environment such as a company officer or personal home, or in a high-risk environment such as a public place, or whether the running application handles potentially sensitive information. These factors can be identified from context information like geolocation, connected/available wifi networks, and running application identification.}, and we use the following abstract function to capture this:
\begin{description}
	\item[$\mathtt{AuthWindow}(c)$] Given context $c$, this algorithm returns an authentication window duration $\Delta t_{window}$.
\end{description}
The fusion algorithm is then implemented by deriving an overall confidence score $\beta$ of the user being present from all individual authentication results obtained within $\Delta t_{window}$.
We will let $\mathcal{H} = \{ \mathcal{H}_{cid}\}_{cid \in \mathcal{S}}$ denote the history of all classifier scores, where $\mathcal{H}_{cid} = \{(\alpha,t)\}$ denotes the list of scores $\alpha$ obtained at time $t$ for each classifier $cid \in \mathcal{S}$, and let $\mathcal{H}_{cid}[t > t']$ denote scores obtained at time $t'$ or later (i.e., $\mathcal{H}_{cid} [t > t'] = \{ (\alpha,t) \in \mathcal{H}_{cid} : t > t'\}$).
As illustrated in Figure~\ref{fig:fusion}, the fusion algorithm will first fuse individual classifier scores via a \emph{sample fusion rule} $\mathtt{SampleFusion}$ to obtain a score $\beta_{cid}$ for each classifier $cid \in \mathcal{S}$; i.e.,\ letting $\Gamma_{cid} \gets \{\alpha : (\alpha,t) \in \mathcal{H}_{cid}[t > t_{now} - \Delta t_{window}] \}$,
\[
\beta_{cid} \gets \mathtt{SampleFusion}(\Gamma_{cid})
\] 
Then an overall score is derived via a \emph{classifier fusion rule} $\mathtt{ClassifierFusion}$:
\[
\beta \gets \mathtt{ClassifierFusion}(\{\beta_{cid}\}_{cid \in \mathcal{S}})
\] 
Since our approach is based on selectively activating classifiers, it is likely that input from only a few classifiers is available. 
On the other hand, as the scheduling algorithm only activates classifiers that are likely to authenticate a genuine user (in the given context), the available scores are likely to be the most reliable scores obtainable from the classifiers.
Hence, in our concrete scheme, we apply a simple approach to $\mathtt{SampleFusion}$ and $\mathtt{ClassifierFusion}$. 
Specifically, after score normalization (e.g., z-score normalization \cite{zscore}), 
we use the average classifier score:
\[
\mathtt{SampleFusion}(\{\alpha\}) = \mathtt{Avg}(\{\alpha\})
\]
where the average of an empty set is defined to be $\bot$, and
\[
\mathtt{ClassifierFusion}(\{\beta_{cid}\}_{cid \in \mathcal{S}}) = \mathtt{Avg}(\{\beta_{cid}\}_{cid \in \mathcal{S}})
\]
where any value $\bot$ is ignored in the average.

The fusion algorithm, using the general $\mathtt{SampleFusion}$ and $\mathtt{ClassifierFusion}$ rules, is shown as Algorithm~\ref{alg:fusion}.

\SetKwInput{KwState}{State}
\begin{algorithm}
\KwIn{$(\mathcal{S},\mathcal{H},c)$ where\\ 
~~~$\mathcal{S} = \{cid\}$: classifier ids;\\
~~~$\mathcal{H} = \{\mathcal{H}_{cid}\}_{cid \in \mathcal{S}}$: history of classifier scores;\\
~~~$c$: context}
\KwOut{Fused score $\beta$}
\BlankLine
	
$t_{now} \gets \mathtt{Clock}()$\;
$\Delta t_{window} \gets \mathtt{AuthWindow}(c)$\;
\tcc{Sample fusion}
$\Gamma \gets \emptyset$\;
\ForEach{$cid \in \mathcal{S}$}{
	$\Gamma_{cid} \gets \emptyset$\;
	\ForEach{$(\alpha_{cid},t) \in \mathcal{H}_{cid}[t > t_{now} - \Delta t_{window}]$}{
		$\Gamma_{cid} \gets \Gamma_{cid} \cup \{\alpha_{cid}\}$\;}
	$\beta_{cid} \gets \mathtt{SampleFusion}(\Gamma_{cid})$\;
	$\Gamma \gets \Gamma \cup \{\beta_{cid}\}$
}
\tcc{Classifier fusion}
$\beta \gets \mathtt{ClassifierFusion}(\Gamma)$\;
\Return{$\beta$}
\caption{Fusion algorithm $\mathtt{Fusion}$.}\label{alg:fusion}
\end{algorithm}

\subsection{Main Authentication Loop}
\label{sec:main_loop}
To realize the framework 
in Figure~\ref{fig:framework-overview}, we use the authentication loop shown in Algorithm~\ref{alg:loop}. 
This continuously obtains the context via the external algorithm $\mathtt{ExtractContext}$ and runs the classifier scheduling algorithm (Algorithm~\ref{alg:sensor_activation}) and fusion algorithm (Algorithm~\ref{alg:fusion}). 
Based on the fused score $\beta$ from the latter, the algorithm updates the device state to be either locked or unlocked via a threshold decision:
\[
	\mathtt{DeviceState} \gets \left\{
	\begin{array}{ll}
		\mathtt{locked} & \text{if } \beta < th_\beta\\
		\mathtt{unlocked} & \text{otherwise }
	\end{array} \right.
\]  
where the threshold $th_\beta$ is a parameter of the framework\footnote{Here, we assume that scores are defined as similarities. 
We can also handle scores defined as distances by locking the device if $\beta > th_\beta$.}.
When the device enters a locked state, the user is required to explicitly authenticate, e.g.,\ via fingerprint scan, PIN, or password, to use the device.

\begin{figure}[t]
	\centering
	\includegraphics[scale=0.25]{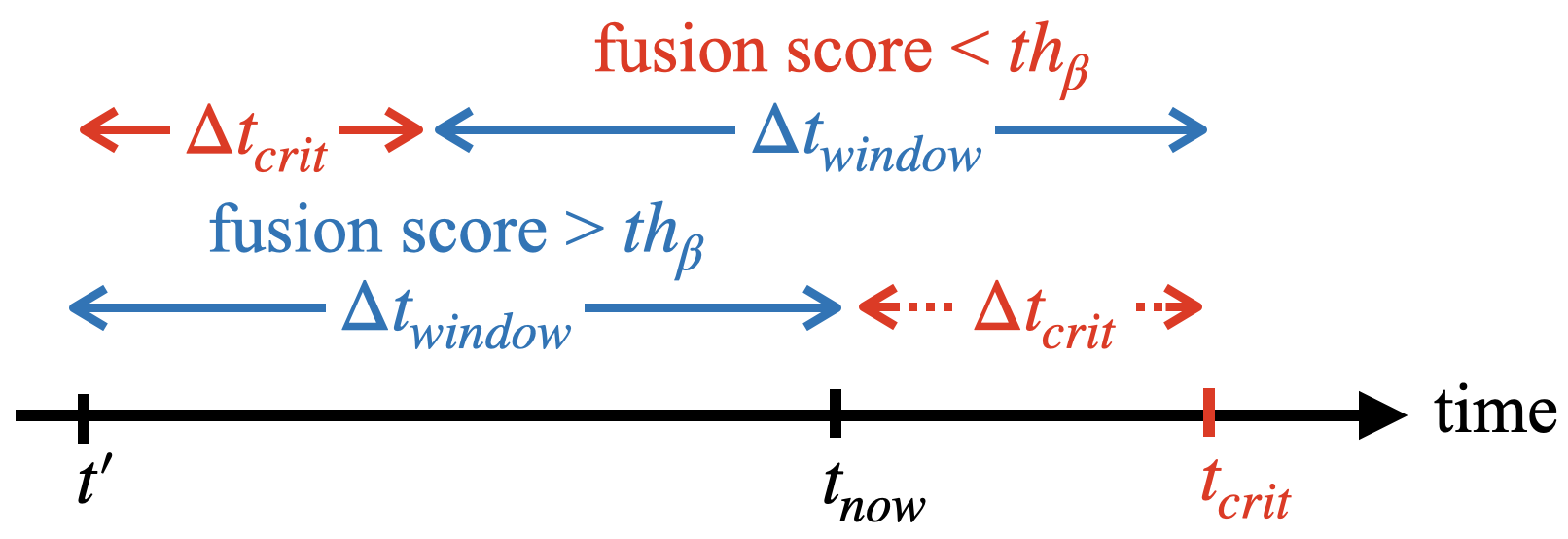}
        \vspace{-2mm}
	\caption{Computation of critical time $t_{crit}$ at which the device will be locked 
    if no classifier scores are obtained from $t_{now}$.
		$t_{crit}$ is determined as the earliest future time where the fusion score, which is computed based on the period of $\Delta t_{window}$ prior to $t_{crit}$, is lower than $th_\beta$. 
    }\label{fig:t-critical}
\end{figure}

The authentication loop maintains the history $\mathcal{H}$ of classifier scores and updates this via the external algorithm $\mathtt{RetrieveLastClassifierResults}$, which is required to return a list of activated classifier results $\mathcal{R} = \{ (cid, \alpha, t)\}$ that have become available since the last invocation, where $cid$ is the classifier id, $\alpha$ the classifier score, and $t$ the time at which the score was obtained.
The history $\mathcal{H}$ is required for running $\mathtt{Fusion}$ and to determine the critical time $t_{crit}$ at which the device is locked unless additional authentication information is available. 
The critical time $t_{crit}$ is determined as the earliest future time where the device will be locked assuming no additional classifier scores are available from $t_{now}$ and
is computed by considering the shortest period of time $\Delta t_{crit}$ the authentication window can be moved forward before the fusion score becomes lower than $th_\beta$; i.e.,\ letting $t' \gets  t_{now} - \Delta t_{window}$, 
\[
\Delta t_{crit} \gets \mathtt{Min}(\{ \Delta t' : \mathtt{Fusion}(\mathcal{H}[t > t' + \Delta t']) < th_\beta \})
\]
where 
$\mathtt{Fusion}(\mathcal{H}[t > t' + \Delta t'])$ denotes a fusion score obtained from scores at time $t' + \Delta t'$ or later.
The critical time $t_{crit}$ is then $t_{crit} = t_{now} + \Delta t_{crit}$; see the illustration in Figure~\ref{fig:t-critical}.
We will denote this computation of $\Delta t_{crit}$ as $\mathtt{CriticalTime}(\mathcal{H},c)$.

After calculating the remaining authentication window $\Delta t_{crit}$, we run the classifier scheduling algorithm (Algorithm~\ref{alg:sensor_activation}) to obtain a set $\mathcal{S}_{act}$ of classifier ids to be activated. 
Then, we activate $\mathcal{S}_{act}$ and calculate scores for them via the external algorithm $\mathtt{ActivateClassifiers}$.

The main authentication loop is shown as Algorithm~\ref{alg:loop}.

\SetKwInput{KwState}{State}
\SetKwInput{KwPar}{Parameters}
\begin{algorithm}
\KwIn{$(\mathcal{S},\mathcal{C})$ where\\ 
~~~$\mathcal{S}$: biometric classifier ids;\\ 
~~~$\mathcal{C}$: context sensor ids.}
\KwPar{$th_\beta$: score threshold for authentication.}
\KwState{History $\mathcal{H} = \{\mathcal{H}_{cid}\}_{cid \in \mathcal{S}}$ of previous classifier scores  $\mathcal{H}_{cid} = \{(\alpha_{cid},t)\}$ for all classifiers $cid \in \mathcal{S}$.}
\BlankLine
\While{True}{
$c \gets \mathtt{ExtractContext}(\mathcal{C})$\;
\tcc{Retrieve activated classifier results $\mathcal{R}$ available since last invocation}
$\mathcal{R} \gets \mathtt{RetrieveLastClassifierResults}(\mathcal{S})$\;
\tcc{Update the history $\mathcal{H}_{cid}$ of scores}
\ForEach{$(cid,\alpha,t) \in \mathcal{R}$}{
	$\mathcal{H}_{cid} \gets \mathcal{H}_{cid} \cup \{(\alpha,t)\}$\;
}
\tcc{Fusion and decision}
$\beta \gets \mathtt{Fusion}(\mathcal{S},\mathcal{H},c)$\;
\uIf{$\beta < th_\beta$}{$\mathtt{DeviceState} \gets \mathtt{Locked}$}
\Else{$\mathtt{DeviceState} \gets \mathtt{Unlocked}$}
\tcc{Classifier activation scheduling}
$\Delta t_{crit} \gets \mathtt{CriticalTime}(\mathcal{H},c)$\;
$\mathcal{S}_{act} \gets \mathtt{ClassifierSchedule}(\mathcal{S},c,\Delta t_{crit})$\;
\tcc{Activate $\mathcal{S}_{act}$}
$\mathtt{ActivateClassifiers}(\mathcal{S}_{act})$\;
$\mathtt{Wait}(\Delta t_{delay})$\;
}
\caption{Main authentication loop.}\label{alg:loop}
\end{algorithm}

\section{Experimental Evaluation}
We evaluate our two-dimensional dynamic fusion via experiments 
using real-world datasets.
In particular, we consider the ability to run our fusion approach at a higher frequency than the existing parallel fusion approaches (e.g., simple max/sum rule \cite{zscore}, CWMA \cite{SivRagSimZic18}). 
For example, suppose three classifiers are available. 
If our approach selects one classifier for each time instant, our approach applied to three time instants calculates the same number of scores as the existing parallel approaches applied to one time instant. 
Thus, we evaluate \textit{how our approach compares with the existing approaches in terms of accuracy with fewer score calculations or with the same number of score calculations}. 
We denote our approach applied to one, two, and three time instants by Our(1x), Our(2x), and Our(3x), respectively. 

\subsection{Uni-Modal Multi-Classifier Experiment}\label{sec:uni-modal-experiment}
\noindent{\textbf{Experimental Setup.}}~~First, we conduct a uni-modal multi-classifier experiment based on data from the extended M2VTS (XM2VTS) database project \cite{xm2vts}.
This is a large dataset containing images of subjects in various poses, 
 including collections of images of subjects facing the camera (`frontal'), subjects in profile (`profile'), 
 as well as images extracted from a movie (MPEG) of the subjects looking in various directions (straight, up, down, left, and right). 
 In the latter images, both eyes of the subject are generally visible, which is not the case for the `profile' images, but the images contain some motion blur. 

From this dataset, we construct two simple classifiers:
\begin{description}
	\item[Classifier 1] based on four `frontal' images for each subject as enrollment images.  
	\item[Classifier 2] based on four `profile' images (two left and two right facing) for each subject as enrollment images.
\end{description}
The classifiers are implemented by using a pre-trained ResNet network with 29 convolution layers to extract a 128-dimensional feature vector for the enrollment images. Likewise, for any test image, a feature vector is extracted and the Euclidian distance to all enrollment images is computed, and finally, the minimum distance is used as a similarity score. 
The scores for both classifiers are normalized by computing the corresponding z-score \cite{zscore}.

We define two contexts for our experiment; a semi-frontal context (SF) based on the MPEG images and a profile context (P) based on the profile images. Specifically, we consider four images per subject (up, down, left, right) for the semi-frontal context and four images (two left, two right, different from enrollment) for the profile context. 
Of the 372 subject IDs in the datasets, 290 have sufficient data for our experimental setup.

For comparison, we consider simple max and sum rules \cite{zscore}, as well as CWMA \cite{SivRagSimZic18}.
The context-dependent weights required for CWMA are obtained by splitting the subject IDs into two sets; one used exclusively for the training of CWMA and the other for the actual experiment. The weights (values in $[0,1]$) are chosen to be the weights that minimize the equal error rate (EER) for the training set found via brute-force computation (precision $0.02$).
Finally, for our approach, the training set is used to estimate the true acceptance rate of the classifiers at the EER point, and a threshold $th_p$ of 
$0.9$ 
was used. 
In this setting, our approach selects one classifier for each time instant\footnote{ 
We use additional separate input (same subject, the same context) when multiple samples are required.}.

\smallskip
\noindent{\textbf{Results.}}~~For the comparison, we 
use detection error trade-off (DET) curves showing the correlation between the 
false acceptance rate (FAR) 
and the  
false rejection rate (FRR). 

\begin{figure*}[t]
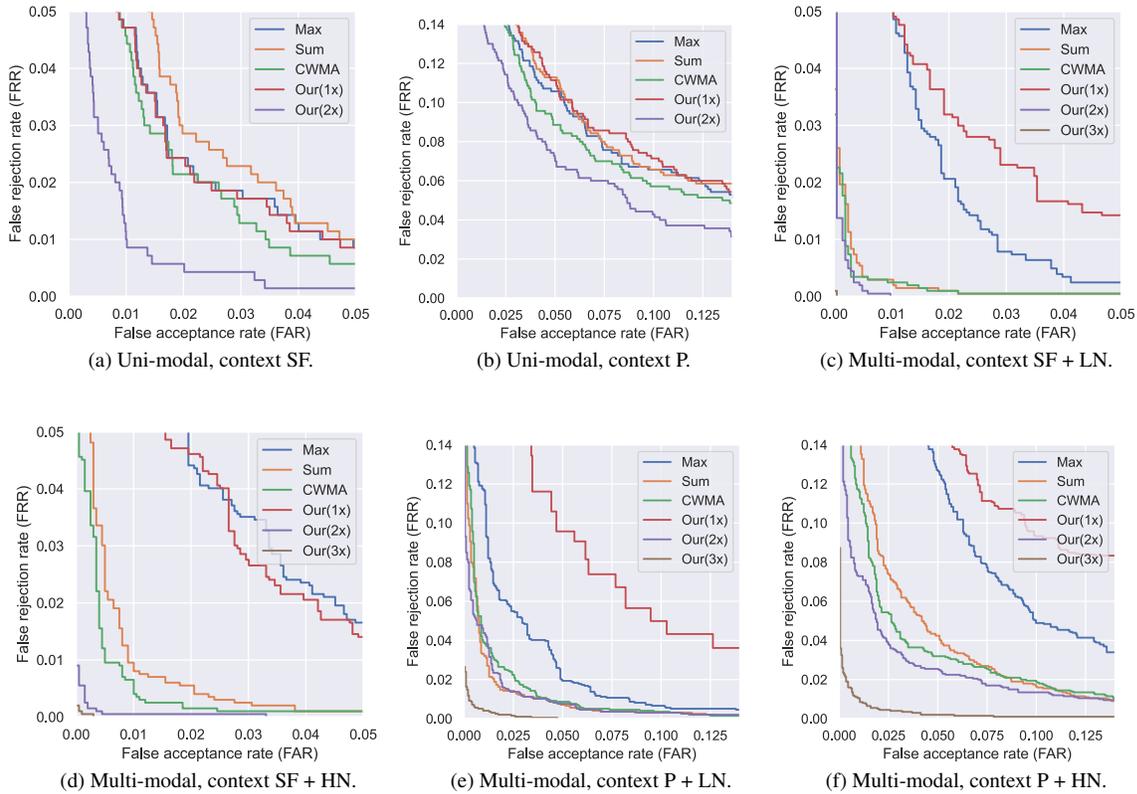

     \centering
     \begin{subfigure}[t]{51mm}
         \centering
         \includesvg[width=\textwidth]{uni-modal-context-0.svg}
         \caption{Uni-modal, context SF.}
         \label{fig:uni-modal-context-0}
     \end{subfigure}
     \begin{subfigure}[t]{49mm}
         \centering
         \includesvg[width=\textwidth]{uni-modal-context-1.svg}
         \caption{Uni-modal, context P.}
         \label{fig:uni-modal-context-1}
     \end{subfigure}
     \begin{subfigure}[t]{51mm}
         \centering
         \includesvg[width=\textwidth]{detonly_contextual_only_ctx11.svg}
         \caption{Multi-modal, context SF + LN.}
         \label{fig:multi-modal-context-11}
     \end{subfigure}\\[4pt]
     \begin{subfigure}[b]{51mm}
         \centering
         \includesvg[width=\textwidth]{detonly_contextual_only_ctx12.svg}
         \caption{Multi-modal, context SF + HN.}
         \label{fig:multi-modal-context-12}
     \end{subfigure}
     \begin{subfigure}[b]{49mm}
         \centering
         \includesvg[width=\textwidth]{detonly_contextual_only_ctx21.svg}
         \caption{Multi-modal, context P + LN.}
         \label{fig:multi-modal-context-21}
     \end{subfigure}
     \begin{subfigure}[b]{49mm}
       \centering
       \includesvg[width=\textwidth]{detonly_contextual_only_ctx22.svg}
       \caption{Multi-modal, context P + HN.}
       \label{fig:multi-modal-context-22}
     \end{subfigure}
     \caption{Detection error tradeoff (DET) curves for uni-modal and multi-modal experiments. Context labels are SF: `semi-frontal'; P: `profile'; LN: `low noise'; and HN: `high noise'.  }\label{fig:two-classifiers-result}
\end{figure*}

Figures~\ref{fig:two-classifiers-result}(a) and \ref{fig:two-classifiers-result}(b) show the DET curves for all of the compared fusion techniques in both of the considered contexts. 
In Tables~\ref{tab:individual-eer} and \ref{tab:fusion-eer}, we also show the EER of each individual classifier and each fusion approach (we explain Classifier 3 and a multi-modal experiment in Section~\ref{sec:multi-modal-experiment}). 
The DET curves demonstrate a clear advantage for our approach when running at a high frequency (i.e.,\ twice as often as the others). Note that assuming an equal cost of the classifiers, this result is achieved without the use of additional resources. 
We also note that in the `semi-frontal' context, where a single classifier greatly outperforms the other, our low-frequency approach (using half the resources compared to the other) performs similarly to the max fusion rule 
(cf. Figure~\ref{fig:two-classifiers-result}(a)), as our approach selects the best classifier.

\begin{table}[t]
\centering
\begin{small}
 \begin{tabular}{|c|c|c|}
		\hline
			Classifier & Context & EER\\
			\hline
		\multirow{2}{*}{\shortstack{Classifier 1\\(`frontal')}} & Semi-frontal & 2.7\%\\
		& Profile & 11.2\%\\
		\hline
		\multirow{2}{*}{\shortstack{Classifier 2\\(`profile')}} & Semi-frontal & 20.4\%\\
		& Profile & 9.2\%\\
		\hline
		\multirow{2}{*}{\shortstack{Classifier 3\\(`voice')}} & Low noise & 7.3\%\\
		& High noise & 17.7\%\\
		\hline		
	\end{tabular}
\end{small}
\caption{EER of individual classifiers for relevant contexts.}\label{tab:individual-eer} 
\end{table}

\begin{table*}[t]
\centering
\begin{small}
	\begin{tabular}{|c|c|c|c|c|c|c|c|c|}
		\hline
		\multirow{3}{*}{Fusion} & \multicolumn{3}{c}{Uni-modal} & \multicolumn{5}{|c|}{Multi-modal}\\
				&  	&  & \#score & &  &  &  & \#score\\
				& SF 	& P & calculations	& SF+LN & SF+HN & P+LN & P+HN & calculations\\
		\hline
		Max 	& 2.2\% & 7.6\% & 2 & 2.1\% & 3.3\% & 4.0\% & 7.5\% & 3\\
		Sum 	& 2.6\% & 7.7\% & 2 & 0.5\% & 1.0\% & 1.7\% & 4.5\% & 3\\
		CWMA 	& 2.1\% & 7.0\% & 2 & 0.3\% & 0.8\% & 2.4\% & 3.6\% & 3\\
               \hline
		Our(1x) & 2.1\% & 8.4\% & 1 & 2.7\% & 2.9\% & 7.4\% & 9.6\% & 1\\
		Our(2x) & 1.0\% & 6.2\% & 2 & 0.3\% & 0.2\% & 1.9\% & 3.2\% & 2\\
		Our(3x) & - 	& -     & -	& 0.1\% & 0.1\% & 0.6\% & 1.1\% & 3\\
		\hline
	\end{tabular}
\end{small}
\caption{%
  Equal Error Rate (EER, lower is better)
  for each fusion approach in uni-modal and multi-modal experiments.
  The number of score calculations are the same across contexts for each approach.
  Context labels are SF: `semi-frontal'; P: `profile'; LN: `low noise'; and HN: `high noise'. %
}\label{tab:fusion-eer}
\end{table*}

\subsection{Multi-Modal Multi-Classifier Experiment}\label{sec:multi-modal-experiment}

\noindent{\textbf{Experimental Setup.}}~~We also conduct a multi-modal multi-classifier experiment by combining the above-described XM2VTS dataset with	the Speakers in the Wild (SITW) Speaker Recognition Challenge dataset \cite{sitw}.
This dataset contains speech samples from open-source media, including samples in noisy and multi-speaker environments.
There are speech samples aimed at training for 50 subjects and evaluation samples for 249 subjects, which are categorized into noise categories.
In our experiment, we only use the evaluation samples in the \verb|core-core| list and separate them into a `low noise' category (SITW categories `clean' and noise levels 0 and 1) and a `high noise' category (SITW categories noise levels 3 and 4).

For the experiment, we consider three classifiers:
\begin{description}
	\item[Classifier 1] identical to Classifier 1 from Section~\ref{sec:uni-modal-experiment}.
	\item[Classifier 2] identical to Classifier 2 from Section~\ref{sec:uni-modal-experiment}.
	\item[Classifier 3] pre-trained voice recognition system, where each subject is enrolled according to the
          \verb|enroll-core| list of the SITW dataset.
\end{description}
The last classifier is implemented via a pre-trained proprietary voice recognition system from FueTrek\footnote{\url{https://www.fuetrek.co.jp/}. Note that this system is not trained on the SITW dataset and is optimized to recognize speech in Japanese of longer length.}.

We define four contexts based on the used data corresponding to the four possible combinations of a semi-frontal or a profile image sample for face recognition, and a low noise (LN) or high noise (HN) speech sample for voice recognition.
Unfortunately, there are very few voice subjects for which we have both a clean and noisy speech sample, and the total number of genuine samples is much lower compared to the possible number of genuine samples for face recognition. 
Hence, we adopt an approach based on random sampling.
Specifically, to sample a genuine sample for a given context, we first randomly pick XM2VTS and SITW subject IDs, with the choice of SITW ID being restricted to the IDs with samples matching the context (i.e., low or high noise). 
Then we randomly pick genuine image and speech samples from all available genuine samples for the chosen IDs and context.
Finally, we run the above three classifiers on the chosen samples to obtain the corresponding scores.    
We take the same approach for all contexts and simply use samples (image and voice) for a different randomly chosen ID to generate impostor samples.
Our dataset contains a total of $152$ and $5800$ genuine speech and face samples, respectively.

As in the uni-modal experiment, we compare our approach to the max and sum fusion rules as well as CWMA, and derive the optimal weights for the latter via brute-force computation on $1000$ samples obtained via the above random sampling process. 
Lastly, for our approach, the classifier performance is determined via $1000$ training samples and a threshold $th_p$ of 
$0.9$ 
is used.
Again, our approach selects one classifier for each time instant in this setting.

\smallskip
\noindent{\textbf{Results.}}~~Figures~\ref{fig:two-classifiers-result}(c) to \ref{fig:two-classifiers-result}(f) show 
the DET curves for all of the compared fusion techniques in all contexts. 

These figures show that 
when running at a medium frequency (2x), our approach performs on par with CWMA and the sum fusion rule 
despite the latter two being based on three classifier inputs. 
In other words, our approach achieves almost the same accuracy as these existing approaches with fewer score calculations. 
Running our approach at a high frequency (3x) yields a clear advantage in all considered contexts, which means that ours achieves higher accuracy with the same number of score calculations.

\subsection{Summary of Results}

The results shown in Sections~\ref{sec:uni-modal-experiment} and \ref{sec:multi-modal-experiment} clearly demonstrates an advantage of our context-aware two-dimensional dynamic fusion approach compared to existing approaches when considering the same number of score calculations. Furthermore, as shown in Section~\ref{sec:multi-modal-experiment}, our approach can achieve similar performance to existing approaches using fewer score calculations, which illustrates the better accuracy and performance trade-off offered by our approach with multi-sample fusion.

Note that the actual resource cost of score calculations may vary between classifiers and will depend on the underlying hardware platform.
Furthermore, some classifier can potentially be optimized for a multi-sample setup\footnote{For example, facial recognition can be combined with face tracking, which is comparable light-weight, to optimize the required image processing and subject identification.} while this might not be possible for other classifiers.
However, overall, the number of score calculations remains a good platform-independent resource measure.

\section{Conclusion}
In this paper, we have proposed a new approach to multi-biometric continuous authentication based on two-dimensional dynamic fusion. In contrast to prior work, our approach dynamically activates classifiers based on context, which enables an optimized multi-classifier and multi-sample fusion.
Our fusion approach 
achieves a better trade-off between accuracy and 
the number of score calculations 
compared to existing approaches, 
which was demonstrated via experimental evaluation using real datasets.

\section*{Acknowledgement}
This research is partly supported by JST CREST Grant Number JPMJCR22M1, JST AIP Acceleration Research JPMJCR22U5 and JSPS KAKENHI Grant Number 19H01109, Japan.

{\small
\bibliographystyle{ijcb2023/ieee}

\begin{thebibliography}{10}\itemsep=-1pt

\bibitem{enterpriseappstoday}
{EnterpriseAppsToday}, {Digital Banking Statistics}.
\newblock
  \url{https://www.enterpriseappstoday.com/stats/digital-banking-statistics.html}.
\newblock Accessed: 2022-12-01.

\bibitem{foxnews}
{Fox News}, {Americans check their smartphones 96 times a day, survey says}.
\newblock
  \url{https://www.q13fox.com/news/americans-check-their-smartphones-96-times-a-day-survey-says}.
\newblock Accessed: 2022-12-01.

\bibitem{lookout}
{Lookout}, {P}hone theft in {A}merica -- {B}reaking down the phone theft
  epidemic.
\newblock
  \url{https://transition.fcc.gov/cgb/events/Lookout-phone-theft-in-america.pdf}.
\newblock Accessed: 2022-12-01.

\bibitem{xm2vts}
{The Extended M2VTS (XM2VTS) Database}.
\newblock \url{http://www.ee.surrey.ac.uk/CVSSP/xm2vtsdb/}.

\bibitem{BBSVN15}
S.~Bharadwaj, H.~S. Bhatt, R.~Singh, M.~Vatsa, and A.~Noore.
\newblock Qfuse: Online learning framework for adaptive biometric system.
\newblock {\em Pattern Recognition}, 48(11):3428--3439, 2015.

\bibitem{CHCBJ15}
D.~Crouse, H.~Han, D.~Chandra, B.~Barbello, and A.~K. Jain.
\newblock Continuous authentication of mobile user: Fusion of face image and
  inertial measurement unit data.
\newblock In {\em 2015 International Conference on Biometrics (ICB)}, pages
  135--142, 2015.

\bibitem{zscore}
A.~Jain, K.~Nandakumar, and A.~Ross.
\newblock Score normalization in multimodal biometric systems.
\newblock {\em Pattern Recognition}, 38(12):2270--2285, 2005.

\bibitem{KPKPS22}
G.~Kalantzis, G.~Papakostas, T.~Karanikiotis, M.~Papamichail, and A.~L.
  Symeonidis.
\newblock A heuristic approach towards continuous implicit authentication.
\newblock In {\em 2022 IEEE International Joint Conference on Biometrics
  (IJCB)}, pages 1--7, 2022.

\bibitem{KPS16}
R.~Kumar, V.~V. Phoha, and A.~Serwadda.
\newblock Continuous authentication of smartphone users by fusing typing,
  swiping, and phone movement patterns.
\newblock In {\em 2016 IEEE 8th International Conference on Biometrics Theory,
  Applications and Systems (BTAS)}, pages 1--8, 2016.

\bibitem{RS22}
S.~Rasnayaka and T.~Sim.
\newblock Action invariant imu-gait for continuous authentication.
\newblock In {\em 2022 IEEE International Joint Conference on Biometrics
  (IJCB)}, pages 1--10, 2022.

\bibitem{Reiner11}
A.~Riener.
\newblock {\em Continuous Authentication based on Biometrics: Data, Models, and
  Metrics}, page~30.
\newblock 09 2011.

\bibitem{Ross06}
A.~Ross, K.~Nandakumar, and A.~K. Jain.
\newblock {\em Handbook of Multibiometrics}.
\newblock Springer, 2006.

\bibitem{RutGab00}
D.~Ruta and B.~Gabrys.
\newblock An overview of classifier fusion methods.
\newblock {\em Computing and Information Systems}, 7:1--10, 01 2000.

\bibitem{SivRagSimZic18}
D.~Sivasankaran, M.~Ragab, T.~Sim, and Y.~Zick.
\newblock Context-aware fusion for continuous biometric authentication.
\newblock In {\em 2018 International Conference on Biometrics (ICB)}, pages
  233--240, 2018.

\bibitem{SR19}
M.~Smith-Creasey and M.~Rajarajan.
\newblock A novel scheme to address the fusion uncertainty in multi-modal
  continuous authentication schemes on mobile devices.
\newblock In {\em 2019 International Conference on Biometrics (ICB)}, pages
  1--8, 2019.

\bibitem{sitw}
{SRI International}.
\newblock {The Speakers in the Wild (SITW) Speaker Recognition Challenge}.
\newblock \url{http://www.speech.sri.com/projects/sitw/}.

\bibitem{VSNR10}
M.~Vatsa, R.~Singh, A.~Noore, and A.~Ross.
\newblock On the dynamic selection of biometric fusion algorithms.
\newblock {\em IEEE Transactions on Information Forensics and Security},
  5(3):470--479, 2010.

\end{thebibliography}

}

\end{document}